\documentclass{PoS}

\title{Chasing extreme blazars with INTEGRAL}

\ShortTitle{Chasing extreme blazars with INTEGRAL}

\author{\speaker{L. Bassani}$^{a}$, M. Molina$^{a}$, R. Landi$^{a}$, A. Malizia$^{a}$, A. J. Bird$^{b}$,
A. Bazzano$^{c}$ and P. Ubertini$^{c}$ \\
\llap{$^{a}$} INAF/IASF Bologna, Italy\\
\llap{$^{b}$} School of Physics and Astronomy, University of Southampton, UK\\
\llap{$^{c}$} INAF/IAPS Rome, Italy\\
E-mail: \email{bassani@iasfbo.inaf.it}}
        
\abstract{Within the blazar population, hard X-ray selected objects are of particular 
interest as they tend to lie at each end of the blazar sequence. In particular, flat spectrum 
radio quasars located at high redshifts display the most powerful jets, the largest black hole 
masses and the most luminous accretion disks: their spectral energy distribution has a Compton 
peak in the sub-MeV region which favours their detection by instruments like INTEGRAL/IBIS and Swift/BAT. 
These sources are even more extreme than blazars selected in other wavebands, like, for example, 
the gamma-ray range explored by Fermi. Here we report on a sample of 12 high redshift blazars 
detected so far by INTEGRAL, including 3 newly identified objects. Some properties of the combined IBIS/BAT sample 
of high redshift blazars (z$\ge$2) are also compared to those of a similar similar sample obtained by Fermi.}

\FullConference{An INTEGRAL view of the high-energy sky (the first 10 years) - 9th INTEGRAL 
Workshop and celebration of the 10th anniversary of the launch\\
		 15-19 October 2012\\
		 Bibliotheque Nationale de France, Paris, France}
		 
\begin{document}

\section{Introduction}
Among Active Galactic Nuclei, the most extreme and powerful objects are blazars.
In the commonly accepted scenario of blazars, a single population of 
high-energy electrons in a relativistic jet radiate from the radio/FIR to the UV/soft X-rays 
through synchrotron radiation and, at higher frequencies, through inverse Compton (IC), scattering 
soft-target photons present either in the jet (synchrotron self-Compton [SSC] model), 
in the surrounding material (external Compton [EC] model), or in both \cite{ghisellini98}. 
In fact, the Spectral Energy Distribution (SED) of 
blazars displays two peaks: the synchrotron one (from the IR to the soft X-rays) and
the Compton peak (MeV/GeV band). The different SEDs observed in blazars
can be explained through the blazar sequence \cite{fossati98}, which proposes that 
the more luminous sources have both synchrotron and Compton 
peaks at lower energies than their fainter (and generally at lower redshifts) counterparts. 
Recently, \cite{ghisellini10} showed that  hard X-ray selected blazars at 
high z are those with the most powerful jets, the most luminous accretion disks and the 
largest black hole masses, i.e. they are even more extreme than blazars selected
in gamma-rays. This can be understood in terms of the blazar sequence,
since the high energy peak in the SED of the most powerful blazars is in the hard X-ray/MeV range. 
As a consequence, these objects are more luminous in the 20--100\,keV band than above 100\,
MeV, and thus become detectable in hard X-ray surveys even if they are undetected 
in gamma-ray ones. Thus, a hard X-ray survey is more efficient in 
finding the most powerful blazars lying at the highest z.
Here we present a sample of 12 high redshift blazars (z$>$1) detected by INTEGRAL,
including three newly reported sources, and discuss their overall properties.

\section{The Blazar Sample}
The sample consists of 12 high redshift blazars: 9 from \cite{malizia12} and 1 from 
\cite{krivonos12}, while 2 have been fully discussed by \cite{bassani12} and \cite{molina12} respectively;
it covers a redshift range from 1 to almost 4.
The table reports the main properties of these objects: redshift, hard and soft X-ray flux, 
black hole mass, 5\,GHz flux as well as detection by either Fermi or BAT or both.   
The sources in the sample are characterised by 20--100\, keV luminosities spanning 
from about 10$^{46}$ to 10$^{48}$ erg\,s$^{-1}$ 
and black hole masses from 12$\times$10$^7$ to at least 5$\times$10$^9$ solar masses;
note also that all have a 5\,GHz radio flux $>$0.1\,Jy. 
Three (IGR J06073--0024, IGR J12319--0749, IGR J17438--0347) of this 12 objects have only recently been identified and in the
following we will summarise their main properties.

\begin{table}
\scriptsize
\centering
\label{sample}
\vspace{0.2cm}
\begin{tabular}{lccccccc}
\hline
{\bf Name}&{\bf z}&{\bf F$_{\rm 20-100keV}^{\ddagger}$}&{\bf F$_{\rm 2-10keV}^{\ddagger}$}&{\bf Mass}               &{\bf F$_{\rm 5GHz}$}&{\bf Fermi/BAT$^{\dagger}$}&{\bf Alt. Name}\\
         &        &                                     &                                 &{\bf (10$^9$M$_{\odot}$)}&{\bf (mJy)}      &                         & \\ 
\hline
Swift J0218.0+7348 & 2.367 & 2.70 & 0.55 & 0.01 & 2278 & Fermi/BAT & S5 0212+73\\
NRAO 140           & 1.258 & 2.46 & 0.72 & 1.8  & 1960 & Fermi/BAT & 4C 32.14 \\ 
PKS 0528+134       & 2.06  & 1.50 & 2.57 & 1    & 1960 & Fermi/BAT & - \\
QSO 0539--2839     & 3.104 & 1.74 & 0.17 & 2    & 1278 & Fermi/BAT & PKS 0537--286 \\
QSO 0836+710       & 2.218 & 5.77 & 2.63 & 3    & 2440 & Fermi/BAT & 4C 71.07 \\
Swfit J1656.3--3302& 2.40  & 3.04 & 0.44 & 2.6  & 290  & Fermi/BAT & - \\
PKS 1830--211      & 2.507 & 4.83 & 1.00 & 1.60 & 5870 & Fermi/BAT & - \\
PKS 2149--306      & 2.345 & 2.51 & 0.80 & 5    & 1720 & Fermi/BAT &  - \\
IGR J22517+2218    & 3.668 & 3.15 & 0.26 & 1    & 174  & -/BAT & -\\
\hline
\multicolumn{8}{c}{Extra Sources}\\
\hline
IGR J06073--0024& 1.08   & $<$3.28 & 0.01 & 0.5 & 96/116 & - & PMN J0606--0025\\
IGR J12319--0749& 3.1 & 0.83 & 0.33 & 2.8 & - & - & - \\
IGR J17438--0347& 1.05 & 0.8 & 1.92 & 3.7 & 2300 & - & PKS 1741--03\\
\hline
\hline
\multicolumn{8}{l}{$^{\ddagger}$: in units of 10$^{-11}$ erg\,cm$^{-2}$\,s$^{-1}$. $^{\dagger}$: source also detected by BAT and/or Fermi.}\\
\end{tabular}
\end{table}

{\bf IGR J06073--0024} is optically classified as a broad emission line AGN (FWHM = 5600\,km\,s$^{-1}$) at z=1.028 
\cite{masetti12}; this source has a counterpart in the CRATES catalogue
with a 8.4\,GHz flux of $\sim$98\,mJy and is also reported in the NVSS survey with a 1.4\,GHz flux
of 133.9\,mJy, yielding to a flat radio spectral index of -0.15.
The statistical quality of the XRT spectrum is such that only an  
estimate of the 2--10\,keV flux is available (F$_{\rm2-10keV}$ = 1.1$\times$10$^{-13}$\,erg\,cm$^2$\,s$^{-1}$). 
At the reported redshift and assuming H$_{\rm 0}$ = 71\,km\,s$^{-1}$\,Mpc$^{-1}$, 
$\Omega_{\Lambda}$ = 0.73 and $\Omega_{\rm M}$=0.27, the source X-ray luminosities are   
3.8$\times$10$^{43}$\,erg\,s$^{-1}$ and 4.1$\times$10$^{45}$\,erg\,s$^{-1}$ 
in the 2--10\,keV and 20--40\,keV band, respectively; this suggests that in IGR J06073--0024
the X-ray spectrum rises towards higher frequencies, as expected in flat spectrum radio quasars (see also 
\cite{molina12} for details).
The mass of the black hole at the centre of this object has been estimated to be of the order of
5$\times$10$^8$M$_{\odot}$ by \cite{masetti12}.

{\bf IGR J12319--0734} is optically classified as a
broad emission line AGN (FWHM = 5600 km s$^{-1}$) at z = 3.1
by \cite{masetti12}; the mass of the central black hole is estimated to be 2.8$\times$10$^9$M$_{\odot}$.
The joint spectral analysis of the XRT and IBIS data \cite{bassani12} indicates that the soft X-
ray spectrum is hard and bright: a simple power law provides an acceptable fit with
$\Gamma$ = 1.35$\pm$0.14. Since in the data to model ratio there is some evidence for a high
energy cut-off (E$_{\rm cut}$), the broad-band spectrum has also been fitted with 
a cut-off power law, yielding $\Gamma$ = 1.24$^{+0.17}_{-0.19}$ and E$_{\rm cut}$ = 24.5$^{+71}_{-15}$\,keV.
Note that the cross-calibration constant between the two instruments, introduced in the fit
to take into account possible flux and/or instrumental mismatches between the two datasets, 
is C = 2.53$^{+5.24}_{-1.55}$, indicating some degree of flux variability. 
Assuming that this broad-band fit represents the average state of the source, 
the estimated observer frame luminosities are 2.8$\times$10$^{47}$ erg\,s$^{-1}$ in the X-ray (2--10\,keV)
band and 7.3$\times$10$^{47}$ erg\,s$^{-1}$ in the 20--100\,keV band, i.e. IGR J12319--0749
has a high energy luminosity similar to those observed in the \cite{ghisellini10} 
sample of high redshifts blazars. The source SED (see \cite{bassani12} for details) resembles that of 
a blazar with the synchrotron peak located between radio 
and near-infrared frequencies and the Compton peak in the hard X-ray band,
as also suggested by the break measured in the combined XRT/IBIS spectrum. 

{\bf IGR J17438--0347}, or PKS 1741--03, \cite{krivonos12} is a quasar at z = 1.054. 
First considered as one of the EGRET sources detected with high confidence \cite{mattox01},
IGR J17438--0347 was not reported in a subsequent re-analysis of
EGRET data \cite{casandjian08} nor has been seen by Fermi so far. It is a strong
(2.3\,Jy at 5\,GHz) radio source with a flat core spectrum,
dominating the radio emission. The source displays variability when monitored
over the 5--22\,GHz range \cite{bach07}. The black hole mass found for
this object is again above 10$^9$M$_{\odot}$ \cite{fan04}.
The X-ray spectrum obtained by the XRT instrument  is
well fitted by a simple power-law with a flat photon index ($\Gamma$$\sim$1.4) and absorbed
by a mild intrinsic column density (N$_{\rm H}$$\sim$0.6$\times$10$^{22}$ cm$^{-2}$). 
It is possible that the lack of counts at low energies is due to absorption
local to the source or to intrinsic spectral curvature, as observed in other
high z blazars \cite{bassani07}.
The 2--10\,keV luminosity is 5.61$\times$10$^{46}$ erg\,s$^{-1}$, while the 20--10\,keV luminosity
is 2.34$\times$10$^{46}$ erg\,s$^{-1}$. 

\section{Conclusions}
So far 17 blazars at redshift higher than 2 have been detected above 10\,keV 
by either {\it INTEGRAL}/IBIS (this work) or {\it Swift}/BAT \cite{baumgartner10} or both; of these, 
10 are above z = 2 and, even more interesting,  7 are above z = 3.
Figures \ref{blazars} compare the sample of distant (z$>$2) blazars detected by Fermi 
with that seen by IBIS and BAT: it is evident that observations in hard X-rays allowed the detection 
of objects at high redshifts (Figure \ref{blazars} left panel) 
and with large black hole masses (Figure \ref{blazars} right panel),
with the same or even better capability than Fermi.
Note also that all the high z blazars so far detected 
at high energies have a 5\,GHz flux $\geq$0.1\,Jy, making this
threshold a possible way of selecting high z blazars in hard X-ray surveys.
As the exploration of the hard X-ray sky continues, it is expected that 
more high redshift blazars with similar characteristics to those presented
here will be found and studied.

\begin{small}
\begin{figure}
\centering
\includegraphics[scale=0.27]{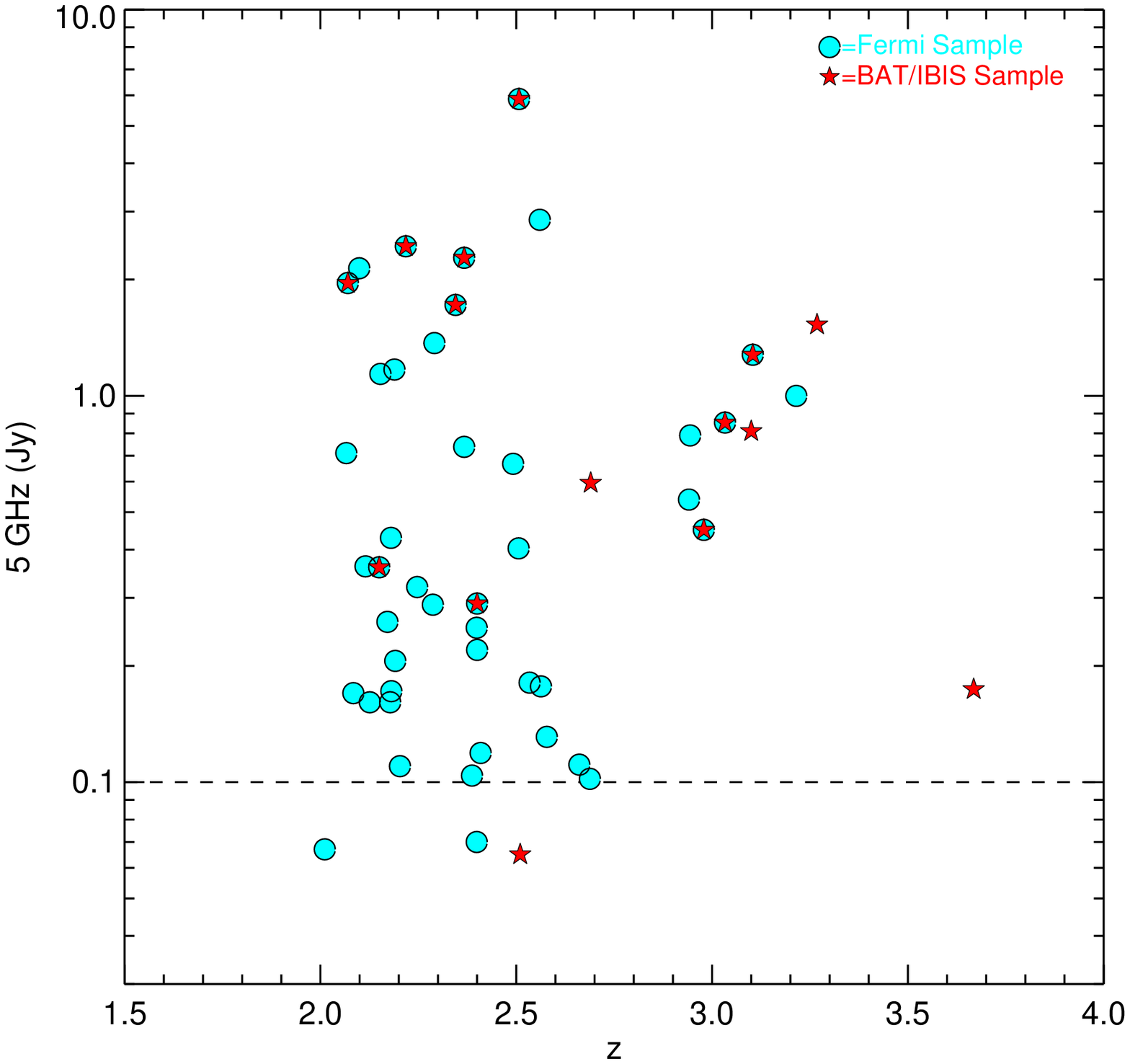}
\includegraphics[scale=0.27]{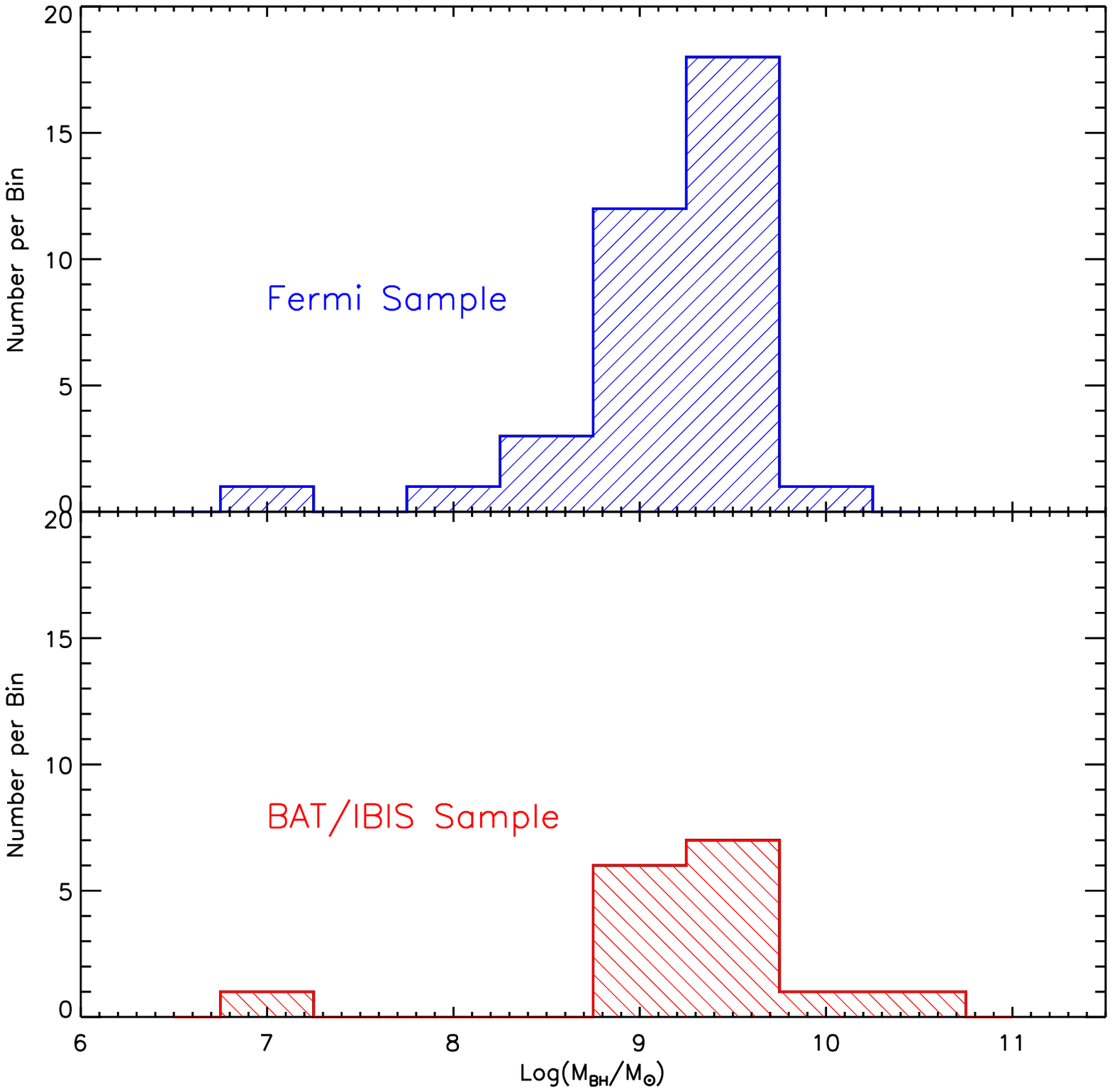}\\
\caption{\small{\emph{Left Panel}: F$_{\rm 5GHz}$ {\it vs.} z for the blazar samples detected by 
Fermi (cyan circles) and by IBIS/BAT (red stars).} 
\emph{Right Panel:} distribution of black hole masses for blazars in the Fermi sample (blue histogram) and 
in the IBIS/BAT sample (red histogram).}
\label{blazars}
\end{figure}
\end{small}

\end{document}